\documentclass[preprint]{aastex}

\begin{document}

\title{Constraining Cosmological Parameters Based on Relative Galaxy Ages}

\author{Raul Jimenez\altaffilmark{1} and Abraham Loeb\altaffilmark{2}}

\altaffiltext{1}{Department of Physics and Astronomy, Rutgers University,
136 Frelinghuysen Road, Piscataway, NJ 08854-8019;
raulj@physics.rutgers.edu}

\altaffiltext{2}{Astronomy Department, Harvard University, 60 Garden
Street, Cambridge, MA 02138; aloeb@cfa.harvard.edu}

\begin{abstract} 

We propose to use relative galaxy ages as a means of constraining
cosmological parameters. By measuring the age difference between two
ensembles of passively--evolving galaxies at somewhat different redshifts,
one could determine the derivative of redshift with respect to cosmic time,
$dz/dt$. At high redshifts, $z\sim 1$--$2$, this measurement would
constrain the equation-of-state of the dark energy, while at low redshifts,
$z\la 0.2$, it would determine the Hubble constant, $H_0$.  The selected
galaxies need to be passively--evolving on a time much longer than their
age difference.

\end{abstract}

\keywords{Cosmology: theory --- galaxies}

\section{Introduction}

Recently, there has been much progress in constraining cosmological
parameters, such as the matter content of the Universe (Peacock et
al. 2001) and the Hubble constant (Freedman et al. 2001).  Available data
for the microwave background anisotropies on degree scales (de Bernardis
et al. 2000; Hanany et al. 2000; Netterfield et al. 2001; Lee et al. 2001 )
and the Hubble diagram of Type Ia supernovae (Riess et al. 1998; Perlmutter
et al.  1999) indicate that the Universe has a flat geometry and is
dominated at present by some form of dark energy with a negative pressure
(Garnavich et al.  1998; Perlmutter et al. 1999). The equation of state of
the dark energy, $p_Q=w_Q \rho_Q$, expresses the ratio between the
pressure, $p_Q$, and the mass density, $\rho_Q$, of the dark energy in
terms of the parameter $w_Q$ (in units of $c=1$).  The value of $w_Q$ could
either be constant as in the case of a cosmological constant ($w_Q=-1$), or
time-dependent as in the case of a rolling scalar field or ``Quintessence''
(Ratra \& Peebles 1988; Caldwell et al. 1998). Any such behavior would have
far--reaching implications for particle physics.  Therefore, the next
observational challenge is to determine the evolution of $w_Q$ as a
function of redshift (Huterer \& Turner 2000; Maor et al. 2001; Weller \&
Albrecht 2001). The related observations need to be done at redshifts $z\la
2$, when the dark energy started to dominate the expansion of the Universe.

The popular approach for measuring $w_Q(z)$ uses its effect on the
luminosity distance of sources.  In particular, the proposal for the
Supernova/Acceleration Probe (SNAP) mission\footnote{http://snap.lbl.gov/}
suggests to monitor $\la 2000$ Type Ia supernovae across the sky per year
and determine their luminosity distances up to a redshift $z\sim 1.5$ with
high precision.  However, the sensitivity of the luminosity distance to the
redshift history of $w_Q(z)$ is compromised by its integral nature (Maor et
al. 2001),
\begin{equation}
d_{\rm L}= (1+z)\int_z^0 (1+z^\prime){dt\over dz^\prime} dz^\prime ,
\label{eq:d_l}
\end{equation}
where $t(z)$ is the age of the Universe at a redshift $z$ which depends on
$w_Q(z)$.

In this paper, we propose an alternative method that offers a much better
sensitivity to $w_Q(z)$ since it measures the integrand of
equation~(\ref{eq:d_l}) directly. Any such method must rely on a clock that
dates the variation in the age of the Universe with redshift.  The clock in
our method is provided by spectroscopic dating of galaxy ages. Based on
measurements of the age difference, $\Delta t$, between two
passively--evolving galaxies that formed at the same time but are separated
by a small redshift interval $\Delta z$, one can infer the derivative,
$(dz/dt)$, from the ratio $(\Delta z/\Delta t)$.  The statistical
significance of the measurement can be improved by selecting fair samples
of passively--evolving galaxies at the two redshifts and by comparing the
upper cut-off in their age distributions. All selected galaxies need to
have similar metallicities and low star formation rates (i.e. a red color),
so that the average age of their stars would far exceed the age difference
between the two galaxy samples, $\Delta t$.

This {\it differential age method} is much more reliable than a method
based on an absolute age determination for galaxies (e.g., Dunlop et
al. 1996; Alcaniz \& Lima 2001; Stockton 2001). As demonstrated in the case
of globular clusters, absolute stellar ages are more vulnerable to
systematic uncertainties than relative ages (Stetson, Vandenberg \& Bolte
1996).  Moreover, absolute galaxy ages can only provide a lower limit to
the age of the Universe and only place weak constraints on the possible
histories of $w_Q(z)$.

The quantity measured in our method is directly related to the Hubble
parameter,
\begin{equation}
H(z)= -{1\over (1+z)}{dz\over dt}.
\label{eq:H_z}
\end{equation}
Hence, an application of this method to old elliptical galaxies in the
local Universe can be used to determine the value of the current Hubble
constant, $H_0$.

In \S 2 we quantify the advantage of the above {\it differential age
method} relative to the standard {\it luminosity distance method} in
constraining the history of $w_Q(z)$.  We then describe the spectroscopic
dating technique and apply it to mock galaxy spectra in \S 3. In
particular, we analyze the dependence of the cosmological constraints on
the signal-to-noise ratio of the spectroscopic data and the number of
observed galaxies. The significance level of the constraints attainable
with a single pair of galaxies, dictates the number of such pairs required
in order to differentiate between various histories of $w_Q(z)$.  Finally,
we summarize our main conclusions in \S 4.

\section{Observables}

We consider a flat universe composed of matter and dark energy, the latter
having an equation of state $p_Q=w_Q(z)\rho_Q$, where $w_Q$ may depend on
redshift. The Hubble parameter is $H^2=H_0^2[\rho_T(z)/\rho_T(0)]$.  Here,
the subscripts $Q$, $m$, and $T$ refer to the dark energy, the matter, or
the total sum of the two, respectively.  Assuming further that the matter
is non-relativistic (i.e. effectively presureless), we get
\begin{equation}
H_0^{-1}{dz\over dt} = - (1+z) {H(z)\over H_0}= -(1+z)^{5/2} \left
[\Omega_m(0) +\Omega_Q(0) \exp \left \{3 \int_{0}^{z}
\frac{dz^\prime}{(1+z^\prime)} w_Q \right \} \right ]^{1/2} ,
\label{eq0}
\end{equation}
where we have used the energy conservation equation for the dark energy,
$\dot \rho_Q=-3H(1+w_Q)\rho_Q$ (Maor et al. 2000).  Thus, $({dz}/{dt})$ is
related to the equation of state of the dark energy through one integration
only, while the luminosity distance in equation (\ref{eq:d_l}) is given by
an integral of the inverse of equation (\ref{eq0}), namely through two
integrations. By differentiating $(dz/dt)$ with respect to $t$ we find,
\begin{equation}
H_0^{-2}\frac{d^2z}{dt^2}=\frac{[H_0^{-1}({dz}/{dt})]^2}{(1+z)}\left
  [\frac{5}{2}+\frac{3}{2}w_Q(z)
  \right]-\frac{3}{2}\Omega_m(0)(1+z)^4w_Q(z) ,
\label{eq2}
\end{equation}
which depends {\it explicitly} on $w_Q$ without any integrations.  Thus,
the second derivative of redshift with respect to cosmic time measures
$w_Q$ directly. While it is possible to find significantly different
redshift histories of $w_Q$ for which the evolution of $d_{\rm L}$ is
similar, this cannot be done for $({d^2z}/{dt^2})$.

Maor et al. (2000) have argued that due to its integral form, the
luminosity distance has only a weak discriminating power with respect to
different possible histories of $w_Q(z)$. To demonstrate their case, they
considered examples of $w_Q$ which have a quadratic dependence on $z$.
Figure 1 shows similar examples (top panel) along with their corresponding
observables, namely the luminosity distance $d_{\rm L}(z)$ (second panel),
${dz}/{dt}$ (third panel), and ${d^2z}/{dt^2}$ (bottom panel).  The
different $w_Q(z)$ histories generate a variation of $\sim 5\%$ in $d_{\rm
L}$, $\sim 10\%$ in $({dz}/{dt})$, and $\sim 30\%$ in $({d^2z}/{dt^2})$ at
$z \sim 1.5$.  The last observable has a fractional variation as large as
that of $w_Q(z)$.

\section{Spectroscopic Dating of Passively--Evolving Gallaxies}

The change of cosmic time with redshift may be inferred from the aging of
stellar populations in galaxies.  This inference must be done with caution
since galaxies are vigorous sites of star formation at high redshift. It
might seem difficult to estimate accurately the differential aging of the
universe based on star--forming galaxies, since the stars in these galaxies
are born continuously and a young stellar population may dominate their
emission spectrum. Fortunately, examples of passively--evolving galaxies
have already been identified in large numbers at high redshifts (Dunlop
1996; Daddi, Cimatti \& Renzini 2000; Dey et al. 2001; Stockton 2001). We
focus our attention on these red galaxies since their light is dominated by
an old stellar population.

Our goal is to measure the age {\it difference} between two
passively--evolving galaxies at different redshifts. {\it How accurately
can this be done?} We emphasize that the relative age is better determined
than the absolute age since systematic effects on the absolute scale are
factored out (for a fractional age difference $\ll 1$). At redshifts $z\sim
1$--$2$ the rest-frame UV spectrum of non star--forming (elliptical)
galaxies is dominated by light from main sequence stars with masses of 1--2
M$_{\odot}$, in a regime of stellar evolution that is well understood
(Spinrad et al. 1997).  The lack of interstellar dust or gas simplifies
further the spectrum of old elliptical galaxies.  The top panel in Figure 2
shows a simulated galaxy spectrum in the rest frame UV that can be easily
collected by 10 meter class telescopes from the ground (e.g. Dunlop et
al. 1996). This spectrum was calculated for a single stellar population
with a specific metallicity (solar) and age (2.4 Gyr), using the synthetic
stellar population models developed in Jimenez et al. (1998).  The
simulated spectrum has a 10 \AA\ resolution, to which we have added Poisson
photon noise with a signal-to-noise ratio $S/N=30$ per resolution
element. The bottom panel shows the values recovered at the 95\% confidence
level (using a $\chi^2$ goodness-of-fit statistics) for the metallicity and
age with $S/N=30$ (dark shaded region) and $S/N=50$ (lightly shaded inner
region). Note that age and metallicity are not degenerate in the
rest--frame UV, provided that the measured spectra have a sufficient $S/N$
(Nolan et al. 2001). A value of $S/N=15$--$30$ can be obtained for a faint
galaxy of magnitude 24 in the $R$--band after 10 hours of integration on
the Keck telescope (K. Adelberger, private communication). Observation of a
sufficiently wide field of view would allow to measure spectra of many
passively--evolving galaxies at the same time.

The assumption that stars in an elliptical galaxy have a single metallicity
value is clearly an over-simplification. Local ellipticals have metallicity
gradients (e.g. Davies et al. 1993), even though these gradients are mild
(e.g. Friaca \& Terlevich 1998). We therefore consider the possibility that
elliptical galaxies have a mixture of populations with different
metallicities\footnote{The small spread in age among stars in an elliptical
galaxy does not affect our results, since the method only relies on the
shift in the {\it average} stellar age as a function of redshift. For a
passively--evolving galaxy, the shift in the {\it average} age of its stars
within a given redshift interval, is determined only by the background
cosmology.}.  In order to assess the accuracy by which we can recover the
age of a mixed stellar population, we simulated several spectra with a
mixture of six metallicity values ranging from 0.01 to 5 times the solar
value. For example, one of our models had weights of 5, 10, 25, 20, 12 and
28\% for metallicities of 0.01, 0.2, 1, 2.5 and 5 times the solar value,
respectively.  We then attempted to recover the age by searching for the
best--fit model with a single metallicity, and found an age uncertainty of
0.1 Gyr. Alternatively, we fitted the simulated spectrum by a mixed
metallicity model with six free components, and recovered a more accurate
age with an uncertainty of $\sim 0.05$ Gyr (as well as a reduced
metallicity uncertainty of 8\%).

Using the more conservative age uncertainty of 0.1 Gyr derived above for
$S/N=30$, we may now estimate the attainable constraints on $w_Q(z)$ for a
fiducial pair of passively--evolving galaxies which formed at the same time
but are observed at $z=1.4$ and 1.6 and have ages of 2.79 and 2.4 Gyr,
respectively.  Figure 3 shows the resulting error bar on $(dz/dt)$, which
is not sufficiently restrictive for a single pair of galaxies.  However,
for a statistical ensemble of independent galaxies (which do not all belong
to the same local region of a galaxy cluster, for example), the error bar
will be reduced by the square root of the number of analyzed galaxy pairs.
The smaller error bar in the plot illustrates the constraint that could be
placed by 20 such pairs. This constraint is sufficient to distinguish among
the different $w_Q(z)$ histories in Figure 1.

The situation in reality is not as simple as described above, since a
galaxy survey will not just identify galaxies that formed at the same
cosmic time.  Rather, the galaxies selected as not having substantial star
formation, will show an age distribution at any given redshift. {\it Given
this age distribution, what is the accuracy by which one may recover the
age shift, provided that the age of each passively--evolving galaxy is
measured with a 0.1 Gyr uncertainty?}  To answer this question, we have
simulated the galaxy age distributions at two different redshifts.  The
distribution at the higher redshift was modeled fiducially as a Schechter
function which is asymptotically flat at low ages and is fully truncated at
some finite maximum age value (because galaxies cannot be older than the
age of the universe). In our example we chose a cutoff at 2.4 Gyr. The
distribution was then binned into bins of 0.1 Gyr width, corresponding to
the age uncertainty for individual galaxies. The age distribution at a
lower redshift was chosen to be the same as the first one but shifted by
either 0.0, 0.15 or 0.30 Gyr (so that the maximum age in the distribution
would be 2.4, 2.55 or 2.7 Gyr, respectively).  We also removed or added a
fixed fraction (20\%) of the galaxies within each bin through a random
Poisson realization of this fraction. This degree of freedom is required
since some galaxies may have merged with a star--forming galaxy and
therefore disappeared from the old galaxy population. Moreover, the
normalization of the two distributions is regarded as a free parameter
because the volume surveyed at the two redshifts might be different.  Our
approach assumes that both samples of galaxies have been passively evolving
for a time much longer than the age difference between the two
redshifts. Each member of the low--redshift sample has a progenitor which,
statistically speaking, is included in the high--redshift sample. But each
high--redshift galaxy has some probability of encountering a merger and
disappearing from the low--redshift sample.  Our method would fail only if
this probability is both significantly large and significantly
age--dependent. However, for a sufficiently narrow redshift interval, the
merger probability is small. The tight color--magnitude relation of
present-day ellipticals further limits the effect of mergers on their age
distribution (Peacock 1991; Bower, Lucey \& Ellis 1992; Peacock et
al. 1998).

The two distributions are shown as the two upper panels of Figure 4 for an
age shift of 0.3 Gyr and a total of $\sim 75$ galaxies in each
distribution. We binned the distributions for illustrative purposes.  The
best statistical method to recover the shift is the Kolmogorov--Smirnov
(KS) test on the unbinned data.  The bottom panel shows the probability for
getting a particular value of the peak in the KS probability (Press et
al. 1992) for the three cases we consider.  We derived this probability
distribution through a large number of Monte--Carlo realizations of the
above distributions.  Also shown as shaded regions are the shift levels for
which one could distinguish among the different $w_Q(z)$ histories in
Figure 3.  We find that the shifts can be recovered with good precision,
despite the fact that the tip of the age distribution is populated by a
small number of galaxies.  In particular, we get a $\ga 90\%$ probability
in this example for distinguishing among the different $w_Q(z)$ histories.

At present, there is no such set of galaxy spectra that could discriminate
between a constant and a variable $w_Q(z)$.  Nevertheless, it is
instructive to examine the best spectroscopically observed pair of red
galaxies to date, namely 53W091 and 53W069, which reside at $z=1.43$ and
1.55, respectively. The stellar populations of these galaxies have been
studied in detail recently by Nolan et al. (2001), who demonstrated that
the age--metallicity degeneracy can be broken based on UV rest-frame
spectroscopy with high quality.  However, since these galaxies where
observed with only $(S/N)\sim 6$--$10$, they do not provide strong
constraints on $w_Q(z)$.

\section{Conclusions}

We have shown that the derivative of cosmic time with respect to redshift
can be used as an effective tool for measuring the equation of state of the
dark energy as a function of redshift. For this purpose, spectra of
passively--evolving galaxies need to be obtained with a high
signal-to-noise ratio.  In order to differentiate at the 2-$\sigma$ level
between a constant and a variable $w_Q(z)$, one needs $\sim 70$ pairs of
galaxies at $R\sim 24$.  For a field of view $\sim (10^\prime)^2$, which is
attainable with instruments like DEIMOS on the Keck telescope and VIMOS on
the VLT, we estimate that the required number of ellipticals may be found
in the volume between $z=1.4$ and 1.6, and so all pairs can be observed
simultaneously over 10 hours of integration on a 10 meter class telescope,
such as Keck or the VLT. 

The key advantages of our method relative to supernova searches such as the
SNAP space mission, are that (i) the observable $(dz/dt)$ is more sensitive
to $w_Q(z)$ than the luminosity distance $d_{\rm L}$; and (ii) the related
precision observations can be done from the ground. In fact, some
observational groups have already identified relevant sets of red galaxies
at high--redshifts (e.g., Daddi et al. 2000; Stockton 2001).

The Hubble constant, $H_0$, can be measured by computing the age difference
between red galaxies at $z=0$ and $z \sim 0.2$. The characteristic age
difference in this redshift range, $\sim 2$ Gyr, can be measured for a
single pair of galaxies with a precision of $\sim 20\%$ given data with
$S/N=10$ and spectral resolution of 10\AA\,. A large volume of
spectroscopic data with this quality on red galaxies in the local universe,
will soon be released by the {\it Sloan Digital Sky Survey}
(SDSS\footnote{http://www.sdss.org}).  The wavelength coverage of the SDSS
spectra of 3900--9100 \AA\, includes spectroscopic features such as
$H_{\beta}$, that are sensitive to the aging of the stellar population.
Given a statistical sample of many hundreds of red SDSS galaxies
(D. Eisenstein, private communication), one could in principle determine
the value of $H_0$ to a percent accuracy, as long as the systematic
modeling uncertainties for relative ages can be reduced to that level.  The
differential age method is independent of the Cepheid distance scale and is
subject to other uncertainties.
 
\acknowledgements

We thank Kurt Adelberger, Rennan Barkana, Rob Caldwell and Marc 
Kamionkowski for useful discussions. This work was supported in part
by NASA grants NAG 5-7039, 5-7768, and by NSF grants AST-9900877, AST-0071019.

\clearpage

\begin{figure}
\plotone{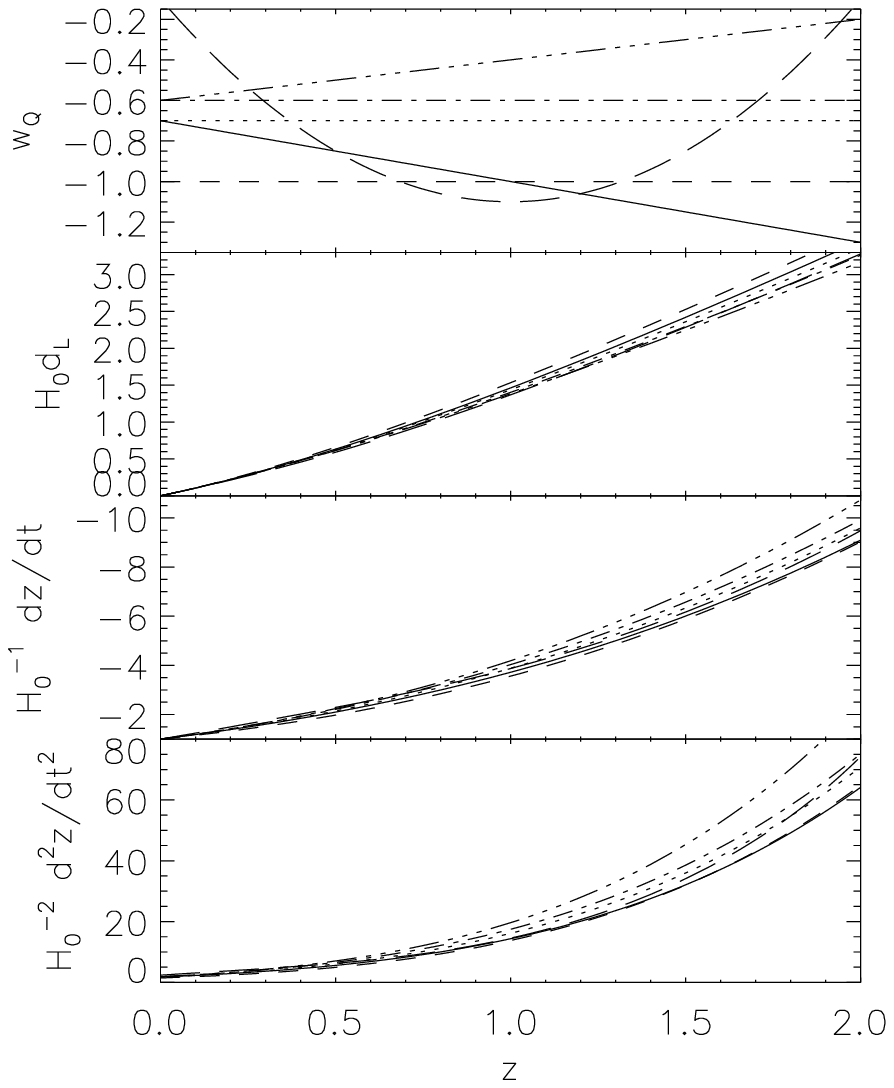} \figcaption{Several examples of $w_Q(z)$ as a
function of redshift (top panel) and the sensitivity of several
observables: $d_{\rm L}$ (second panel), $dz/dt$ (third panel) and
$d^2z/dt^2$ (bottom panel).}
\label{fig1}
\end{figure}

\clearpage
\begin{figure}
\plotone{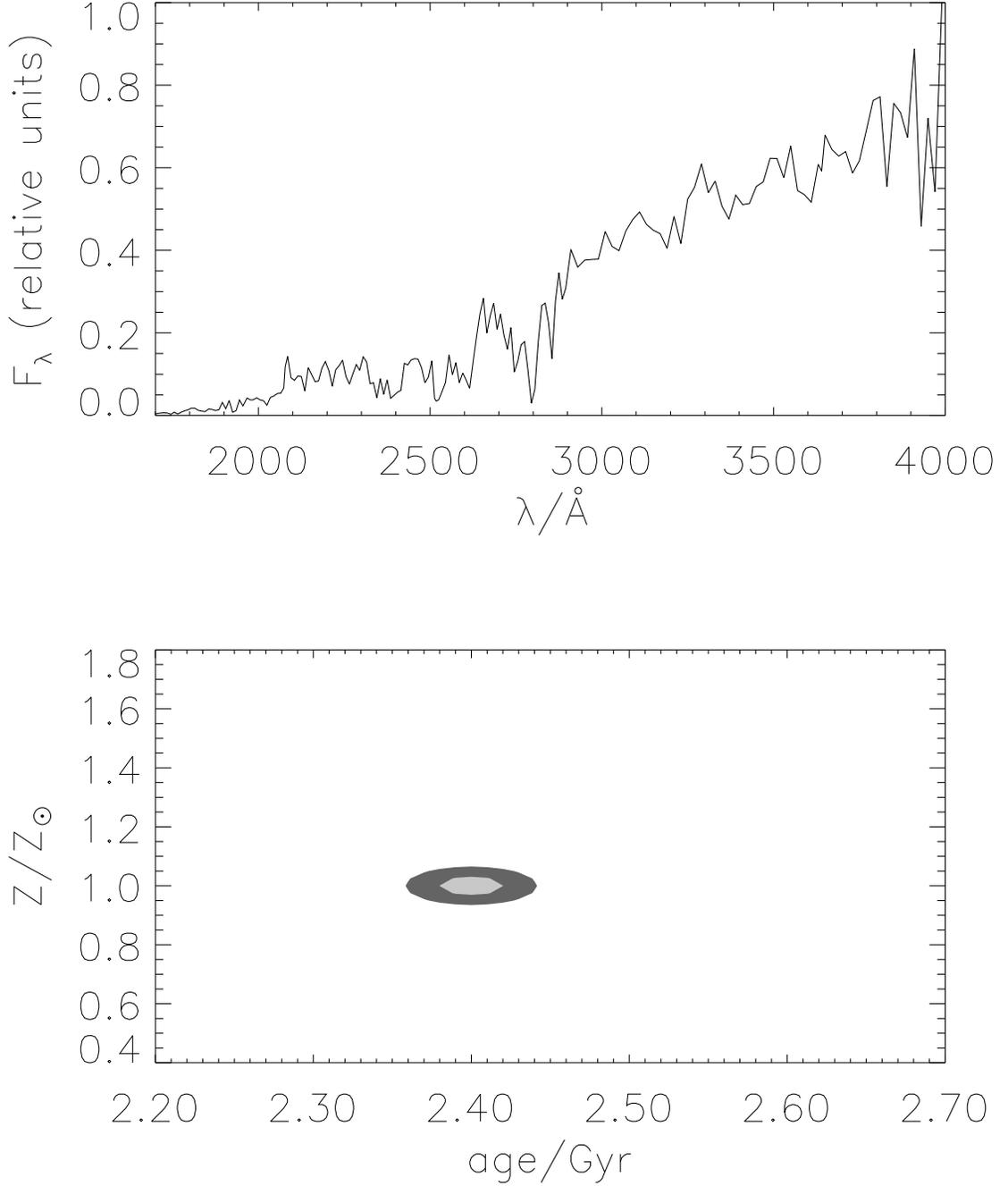} \figcaption{The top panel shows a simulated spectrum
(flux per unit wavelength, $F_\lambda$) for a galaxy with a solar
metallicity and an age of 2.4 Gyr, including Poisson photon noise with
$S/N=30$ at a spectral resolution of 10\AA.  The bottom panel shows the
accuracy of the metallicity and age recovery for $S/N=30$ (dark shaded
region) and $S/N=50$ (lightly shaded region).}
\label{fig2}
\end{figure}

\clearpage

\begin{figure}
\plotone{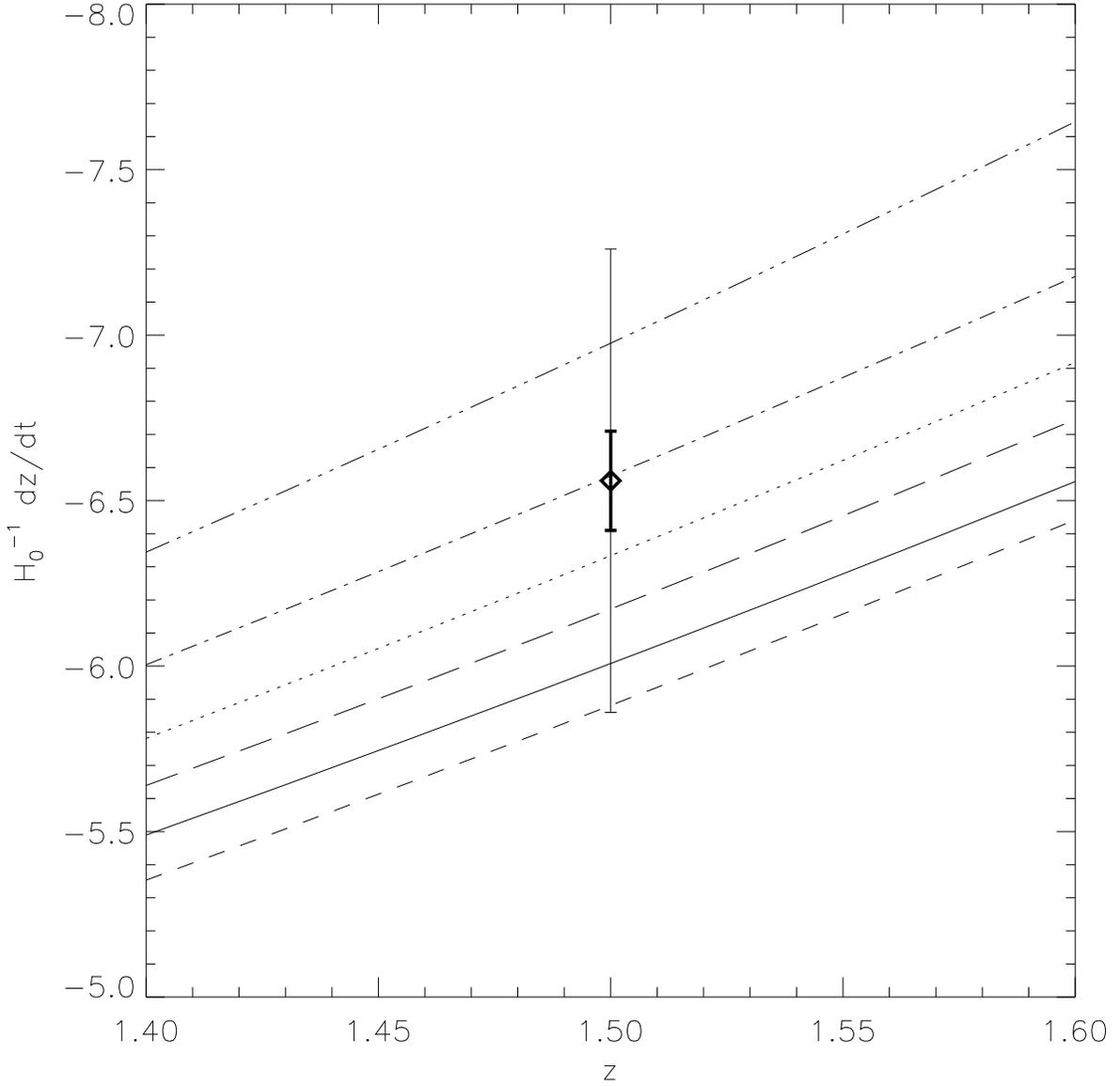} \figcaption{The large error bar illustrates the
constraint that can be placed on $w_Q(z)$ by using only a single pair of
red galaxies that formed at the same time and are observed at $z=1.6$ and
$z=1.4$. The spectra are assumed to have a spectral resolution of 10\AA\,
and $S/N=30$. When data for 20 such pairs is available, the uncertainty is
reduced to the thicker, smaller error bar.}
\label{fig3}
\end{figure}

\clearpage

\begin{figure} 
\plotone{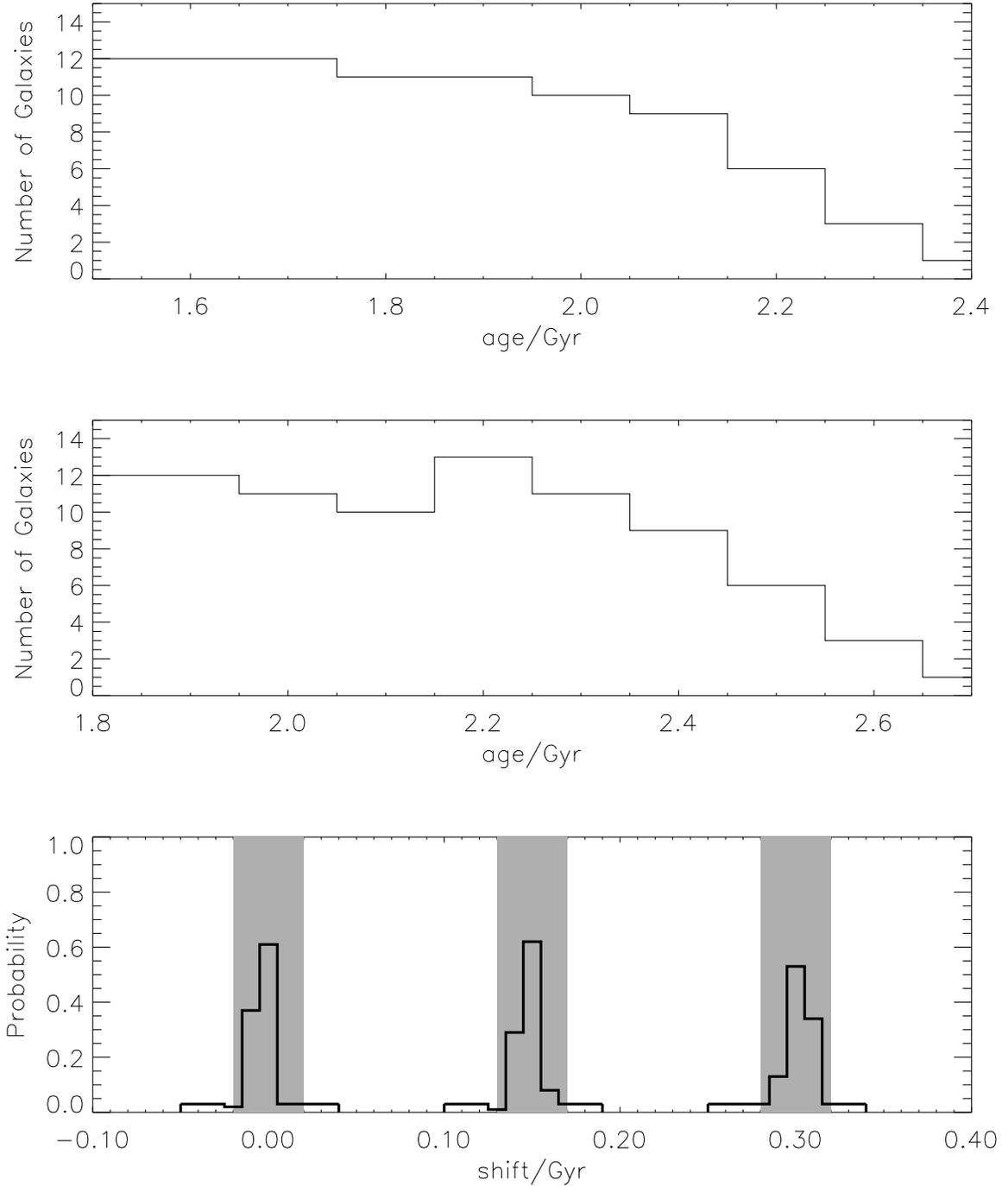}
\figcaption{The top panel shows a hypothetical age distribution for 75
galaxies at a redshift $z=1.6$. The middle panel shows the same
distribution at $z=1.4$, in which all galaxies are older by 0.3 Gyr and
their fraction per bin changes by $\pm 20\%$ with Poisson fluctuations (due
to mergers).  The bottom panel shows the probability distribution of the
recovered shift in age between the two galaxy distributions using the peak
probability value of the KS--likelihood test for unbinned data with an age
error of 0.1 Gyr per galaxy. We show three cases of shifts, namely: 0.0,
0.15 and 0.3 Gyr.  The shaded regions show the range in the recovered shift
which would allow to separate the different $w_Q(z)$ histories in Figure
3.}
\label{fig4} 
\end{figure}

\begin{thebibliography}{}

\bibitem[Alcaniz \& Lima(2001)]{2001ApJ...550L.133A} Alcaniz, J.\ S.\ \& 
Lima, J.\ A.\ S.\ 2001, \apjl, 550, L133 


\bibitem[]{} Bower, R.\ G., Lucey, J.\ R., Ellis R.\ S. 1992, MNRAS, 254, 601

\bibitem[Caldwell, Dave, \& Steinhardt(1998)]{1998Ap&SS.261..303C}
Caldwell, R.\ R., Dave, R., \& Steinhardt, P.\ J.\ 1998, \apss, 261, 303

\bibitem[de Bernardis et al.(2000)]{2000Natur.404..955D} de Bernardis, P.\ 
et al.\ 2000, \nat, 404, 955 

\bibitem[]{} Daddi, E., Cimatti, A., Renzini, A.\ 2000, A\&A, 362, 45

\bibitem[]{} Dey et al.\ 2001, in preparation

\bibitem[Dunlop et al.(1996)]{1996Natur.381..581D} Dunlop, J., Peacock, J.,
Spinrad, H., Dey, A., Jimenez, R., Stern, D., \& Windhorst, R.\ 1996, \nat,
381, 581

\bibitem[Freedman et al.(2000)]{2000astro.ph.12376F} Freedman, W.\ L.\ et
al.\ 2001, ApJ, in press; astro-ph/0012376

\bibitem[Garnavich et al.(1998)]{1998ApJ...509...74G} Garnavich, P.\ M.\ et
al.\ 1998, \apj, 509, 74

\bibitem[Hanany et al.(2000)]{2000ApJ...545L...5H} Hanany, S.\ et al.\ 
2000, \apjl, 545, L5 

\bibitem[Huterer \& Turner(2000)]{2000astro.ph.12510H} Huterer, D.\ \&
Turner, M.\ S.\ 2000, Phys.\ Rev.\ D, submitted; astro-ph/0012510

\bibitem[]{} Jimenez, R.\, Padoan, P.\, Matteucci, F. \, \& Heavens, A. \ F. \
  1998, MNRAS, 299, 123 

\bibitem[]{} Lee, A.T.\ et al.\ 2001; astro-ph/0104459

\bibitem[]{} Maor, I., Brustein, R., \& Steinhardt, P. J.  2001,
Phys. Rev. Lett. 86, 6

\bibitem[]{} Netterfield, B.\ et al.\ 2001; astro-ph/0104460

\bibitem[]{} Nolan, L. A., Dunlop, J. S., Jimenez, R., \& Heavens, A. F.
2001, MNRAS, submitted; astro-ph/0103450

\bibitem[]{} Peacock, J.\ A. 1991, in Blanchard A., Celnekier L.,
  Lachieze--Rey, M., Tran Thanh Van J., eds, Proc. 2nd Rencontre de Blois,
  Physical Cosmology. Editions Frontieres, Gif-sur Yvette, p.337

\bibitem[]{} Peacock, J. A. et al.\ 1998, MNRAS, 296, 1089

\bibitem[]{} Peacock, J. A. et al.\ 2001; astro-ph/0103143

\bibitem[Perlmutter et al.(1999)]{1999ApJ...517..565P} Perlmutter, S.\ et 
al.\ 1999, \apj, 517, 565 

\bibitem[Perlmutter, Turner, \& White(1999)]{1999PhRvL..83..670P} 
Perlmutter, S., Turner, M.\ S., \& White, M.\ 1999, Physical Review 
Letters, 83, 670 

\bibitem[]{} Press, W.\ H., Teukolsky, S. \ A., Vetterling, W. \ T., Flannery,
  B. \ P. 1992, Numerical Recipes. The art of scientific computing. Cambridge
  University Press.

\bibitem[Ratra \& Peebles(1988)]{1988PhRvD..37.3406R} Ratra, B.\ \& 
Peebles, P.\ J.\ E.\ 1988, \prd, 37, 3406 

\bibitem[Riess et al.(1998)]{1998AJ....116.1009R} Riess, A.\ G.\ et al.\ 
1998, \aj, 116, 1009 

\bibitem[]{} Spinrad, H.\ et al.\ 1997, \apj, 484, 581

\bibitem[]{} Stetson, P. \ B., Vandenberg, D. \ A., \& Bolte, M. \, 1996, PASP,
  108, 560

\bibitem[]{} Stockton, A. 2001, to be published in Astrophysical Ages and
Time Scales, ASP Conference Series; astro-ph/0104191

\bibitem[]{} Weller, J. \& Albrecht, A. 2001, Phys. Rev. Lett., 86, 1939
\end{thebibliography}
\end{document}